\documentclass[12pt,preprint]{aastex}

\begin{document}

\title{Variable Infrared Emission from the 
Supermassive Black Hole at the Center of the Milky Way}
\author{
A. M. Ghez\altaffilmark{1,2}, 
S. A. Wright\altaffilmark{1}, 
K. Matthews\altaffilmark{3}, 
D. Thompson\altaffilmark{3}, 
D. Le Mignant\altaffilmark{4}, 
A. Tanner\altaffilmark{1}, 
S. D. Hornstein\altaffilmark{1}, 
M. Morris\altaffilmark{1}, 
E. E. Becklin\altaffilmark{1}, 
B. T. Soifer\altaffilmark{3}}

\altaffiltext{1}{UCLA Department of Physics and Astronomy, Los Angeles, CA 90095-1562; ghez, saw, tanner, seth, morris, becklin@astro.ucla.edu}
\altaffiltext{2}{UCLA Institute of Geophysics and Planetary Physics, 
Los Angeles, CA 90095-1565}
\altaffiltext{3}{Caltech Optical Observatories, California Institute of Technology, MS 320-47, Pasadena, CA 91125; kym, djt, bts@irastro.caltech.edu}
\altaffiltext{4}{W. M. Keck Observatory, 65-1120 Mamalahoa Hwy, Kamuela, HI 96743; davidl@keck.hawaii.edu}

\begin{abstract}

We report the detection of a variable point source, imaged 
at L'(3.8 $\mu$m) with the Keck II 10 m telescope's adaptive optics 
system, that is coincident to within 18 mas (1 $\sigma$)
of the Galaxy's central supermassive black hole and the unique radio source
Sgr A*.  While in 2002 this source (SgrA*-IR) was confused with the stellar 
source S0-2, in 2003 these two sources are separated by 87 mas 
allowing the new source's properties to be determined directly.
On four separate nights, its observed L' magnitude ranges from 
12.2 to 13.8, which corresponds
to a dereddened flux density of 4 - 17 mJy;
no other source in this region shows such large variations
in flux density - a factor of 4 over a week and a factor of 2 over 40 min.
In addition, it has a K-L' color greater than 2.1, which is at least 
1 mag redder than any other 
source detected at L' in its vicinity.  
Based on this source's coincidence
with the Galaxy's dynamical center, its lack of motion, 
its variability, 
and its red color, we conclude that it is associated with the central 
supermassive black hole.
The short timescale for the 3.8 $\mu$m flux density variations 
implies that the emission arises quite close to the black hole, within
5 AU, or 80 R$_{s}$.
We suggest that both the variable 
3.8 $\mu$m emission and the X-ray flares
arise from the same underlying physical process, possibly the 
acceleration of a small populations of electrons to ultrarelativistic
energies.  In contrast to the X-ray flares which are 
only detectable $\sim$2\% of the time, 
the 3.8 $\mu$m emission provides a new, constantly accessible,
window into the physical
conditions of the plasma in close proximity to the central black hole.
\end{abstract}


\section{Introduction} \label{sec_intro}

At the center of the Milky Way lurks a supermassive black hole (SMBH), 
whose presence and mass, $\sim$4$\times 10^6 M_{\odot}$, has been 
ferreted out through its strong 
gravitational field (e.g., 
Ghez et al. 2003a,b; Sch\"odel et al. 2002, 2003), but
whose associated radiative emission has remained more elusive.
For decades it was only detected at radio wavelengths (see, e.g., 
Melia \& Falcke 2001), where it appears to show low amplitude intensity
variations with a quasi-periodicity of $\sim$ 106 d 
(Zhao, Bower, \& Goss 2001).  
Recent observations detected it at X-ray wavelengths, where
it appears to 
be composed of the following two distinct components 
(Baganoff et al. 2001, 2003a,b;
Goldwurm et al. 2003; Porquet et al. 2003):
(1) a steady state, which has been stable to within $\sim$10\% 
over the past 4 years and which is spatially resolved with a size that 
corresponds to the Bondi radius ($\sim$1$\tt''$), and (2) an unresolved 
variable component, which has a flux density that rises above the quiescent
level by an order of magnitude for $\sim$ 1 hour approximately once a day.
These detections at opposite ends of the electromagnetic spectrum, 
along with existing infrared (IR) limits (e.g., 
Serabyn et al. 1997; Morris et al. 2001; Hornstein et al. 2002) reveal that overall, Sgr A* is 
remarkably weak, with a bolometric luminosity of only 10$^{36}$ ergs s$^{-1}$ 
or, equivalently, $10^{-9} L_{Edd}$. 

Detection of emission at IR wavelengths would be a very powerful
constraint for the proposed models of the accretion flow onto the 
SMBH as well as possible outflows (e.g., 
Markoff et al. 2001; 
Liu \& Melia 2002; Yuan et al. 2003a).  At mid-IR wavelengths ($\gtrsim$8 $\mu$m), the 
detectability of a source is limited by thermal emission from dust
(Stolovy et al. 1996; Morris et al. 2001).  At near-IR wavelengths
(1-2 $\mu$m), detection of an associated point source is 
difficult due to the confusion with stellar sources 
(e.g., Close et al. 1995; Genzel et al. 1997; 
Hornstein et al. 2002).  In the 3-5 $\mu$m window,
the stellar and dust emission are both declining, potentially
allowing an unambiguous detection of a radiative counterpart at a wavelength
intermediate to the radio and X-ray regimes. 

In this paper, we present new L'(3.8 $\mu$m) images from the W. M. Keck II 
10-meter telescope of the Galactic center showing a near-IR
counterpart to Sgr A*.
Section 2 describes the 
observations and \S 3 summarizes the data analysis.
Section 4 reports
the detection of a new L' source and its properties.  Finally, 
\S 5 discusses possible emission mechanisms for the new source and concludes
that it is indeed associated with the SMBH.

\section{Observations} \label{sec_obs}

Adaptive optics L' ($\lambda_o$ = 3.8 $\mu$m \& $\Delta \lambda$ = 
0.7 $\mu$m) bandpass 
observations of the Galactic center were obtained
with the W. M. Keck II 10 m telescope using the facility near-infrared
camera, NIRC2 (Matthews et al., in prep) on the following 4 nights:
2002 May 31, 2003 June 10, 16, \& 17 (UT).
The positions of IRS 16NE, NW, \& SW with respect to
IRS 16C in the final map (see \S3) compared to that reported in Ghez et al.
(2003b) established a plate scale of 
9.93 $\pm$ 0.05 mas pix$^{-1}$,
corresponding to a field of view of 10\farcs 2 
$\times$ 10\farcs 2 for NIRC2's narrow field camera.
Using an R=13.2 mag natural guide star located 30$\tt''$ from Sgr A*,
we were able to achieve a resolution as high as 80 mas, which corresponds 
to the diffraction limit,  and Strehl ratios as high as $\sim$0.4.
Each night, several images, composed of 0.2 to 0.5 sec coadded 
exposures with effective exposure on the sky of 20 to 60 sec,
were collected using a 5 position dither pattern with offsets of
$\sim$1 arcsec.  

\section{Data Analysis} \label{sec_anal}

The standard image reduction steps of sky subtraction, flat fielding,
and bad pixel removal were carried out on each image.
Visual inspection of the reduced individual images eliminated those that were 
degraded by poor adaptive optics tip-tilt corrections, which occasionally 
generated double peaked PSFs.  Inclusion in the final map
for the remaining images was based on  their estimated
Strehl ratios, which were required to be $\gtrsim$0.2. 
All of the selected images were registered and 
averaged together using the centroid of IRS 16C's light distribution.  

Sources were identified and their positions and relative intensities were 
quantified using the point spread function (PSF) fitting program StarFinder
(Diolaiti et al. 2000).  Three bright point sources (IRS 16C, 16NW, \& 33N) 
were entered into StarFinder's algorithm for generating a model 
PSF.  
Uncertainties in this process were established by dividing the data set up
into independent maps composed of 2 to 3 frames, 
re-running this procedure on each map, and
taking root-mean-square values for the astrometric and relative 
photometric values.  This also allowed an examination of the data set 
for shorter timescale flux density variations.

Measurements reported here are photometrically calibrated based
on absolute L' photometry for several bright stars within our fied of view
obtained by Simons \& Becklin (1996).
From the 8 stars in common, we selected the
following 4 stars, which are not known to 
be variable or spatially extended at L' or at K (2.2 $\mu$m), as
calibration sources: IRS 16NE (7.77 mag), IRS 16SW-E (8.30 mag),
IRS 16NW (8.87 mag), \& IRS 16C (also labeled IRS 16C-W; 8.91 mag).  
Our multiple measurements of these sources reveal 
no evidence for variability, confirming their suitability as local standards.
We estimate the uncertainty
in the apparent magnitude calibration for each image to be 0.1 mag from the 
standard deviation of the mean of the calibration sources.
All the reported measurements are based on a zero point 
of 248 Jy for the L' magnitude scale (Tokunaga 2000).

The inferred position of the SMBH in the L' maps is based on its dynamically 
determined position.  Using the orbital solutions 
for S0-1, S0-2, S0-3 and S0-4 reported in Ghez et al. (2003b), we localize the 
SMBH in each of the L' images to within $\pm$ 5-10 mas, 
where these uncertainties are limited by inaccuracies in the centroids of 
the stars in the L' maps.

\section{Results}

\begin{figure}
\epsscale{1.0}
\plotone{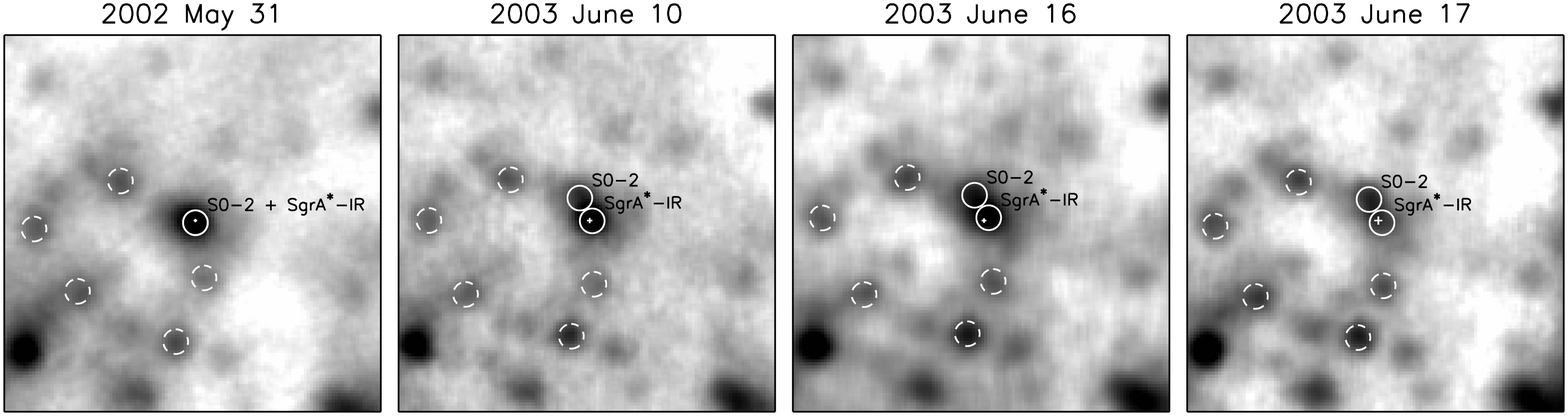}
\figcaption{
A 1.2$\tt''$ $\times$ 1.2$\tt''$ region of the L'(3.8 $\mu$m) Galactic center 
images obtained with the adaptive optics systems on the W. M. Keck II 10-meter 
telescope during the nights of 2002 May 31, 
2003 June 10, 2003 June 16, and 2003 June 17.  In each image, a cross denotes the dynamically determined 
position of the central black hole 
and its uncertainties.  An L' source, SgrA*-IR, is coincident
with this position in all images; while in 2002 SgrA*-IR is blended with 
the stellar source S0-2, S0-2's significant orbital motion makes it well 
resolved from SgrA*-IR in the 2003 images.
Clear intensity variations are detectable in these nightly images, in which
SgrA*-IR is at its brightest in 2003 June 10, when it dominates over S0-2, 
and its faintest in 2003 June 17, when it is somewhat dimmer than S0-2.
The 5 comparison stars, whose photometry is plotted in Figure 2, are circled
with dashed lines.
\label{fig_lp}}
\end{figure}

Figure~\ref{fig_lp} shows a 1\farcs 2 $\times$ 1\farcs 2 region of the combined 
images centered on the SMBH's location for each of the four nights
reported here.  In 2003, a new source is detected.  With the exception 
of this new source all other 
detected L' sources in Figure~\ref{fig_lp} are coincident with
stars detected at 2.2 $\mu$m (Ghez et al. 2003b).
The new source distinguishes itself in 
several ways.  It is an unresolved point source that has 
a negligible average offset of 12 $\pm$ 6 mas from the SMBH. 
In contrast to stars in its vicinity, the 
new L' source is remarkably variable.  This is shown in 
Figure~\ref{fig_phot}, which plots the new source's dereddened flux density 
along with that of all sources that are nearby
(r $<$ 0\farcs 5), are comparably bright (L' $\lesssim$ 13.6 mag), and do not 
suffer from confusion; all the sources are dereddened assuming an L'
extinction of 1.83 mag, which is based on a visual
extinction of 30 mag from Rieke, Rieke, \& Paul (1989) and an extinction law
from Moneti et al. (2001).  On 2003 June 10 the new source was at its 
brightest at 12.28 mag and then it dimmed by a factor of 3 (13 $\sigma$) 
to 13.38 mag 
by 2003 June 17, with a factor of 2 (12 $\sigma$) change occurring between June 16 and 17.   
Subdivision of the data into shorter time blocks of 60 second exposures,
which correspond to elapsed time of 100 - 250 seconds depending on how images
were selected (see \S3), shows significant substructure (see Figure~\ref{fig_phot});
on 2003 June 17 the new source is observed to have dimmed by a factor of 2, with
a significance of 5 $\sigma$.
This also increases the range of intensity variations observed over a week
to a factor of 4.
Table~\ref{tab_sum} provides a summary of this source's 
measured properties, which includes its position with
respect to the SMBH as well as its apparent magnitude 
and dereddened flux density at 3.8 $\mu$m. 

In our 2002 map, the new L' source is blended with the star S0-2, which at 
this time was experiencing its closest approach to the SMBH 
with a projected separation of a mere 14 mas 
2003a,b).
This confused the 
identification of the L' source in this first epoch map, as well
as those obtained at the VLT on 2002 August 29 (Cl\'enet et al. 2003, Genzel
et al. 2003)\footnote{
There is a significant zero-point offset
between our reported values and what is presented by Cl\'enet et al. (2003) and
Genzel et al. (2003); using stellar sources in common within the central
0\farcs 7, we measure the offsets to be L'$_{Keck}$ = L'$_{Genzel}$ + 0.82 mag
\& L'$_{Keck}$ = L'$_{Clenet}$ + 0.37 mag.  This is primarily a consequence 
of a different choice of calibration sources, which includes known variable
and extended stars in their sample
(see \S3); 
Cl\'enet et al. (2003)
use IRS 16C, 29N, 16CC, 21, 33SE, \& MPE+1.6-6.8 and Genzel et al. (2003) 
restrict their analysis of the same data set to IRS 16CC \& 33N. 
}
In our 2003 maps, S0-2 is 87 mas away from the dynamical center and is
clearly separated from the newly identified L' source (see Figure~\ref{fig_lp}). 
Since S0-2 shows no evidence for significant variability in the K-band, 
where there is
negligible confusion with other sources (Ghez et al. 2003b),
we use the L' mag of S0-2 measured in 
2003 June 16 \& 17,  when the contrast between S0-2 and the new source is 
smallest, ($<$L'$_{S0-2}>$ = 13.07) to subtract the S0-2 contamination 
from
the new source in 2002 May 31.
This results in an inferred brightness of L'=13.2 mag, which is
close to the average of the values measured in 2003 June.  
A comparison of the 2002 position with the average 2003 position 
limits its proper motion to 400 $\pm$ 300 km s$^{-1}$.

\section{Discussion \& Conclusions} \label{sec_disc}

\begin{figure}
\epsscale{1.0}
\plotone{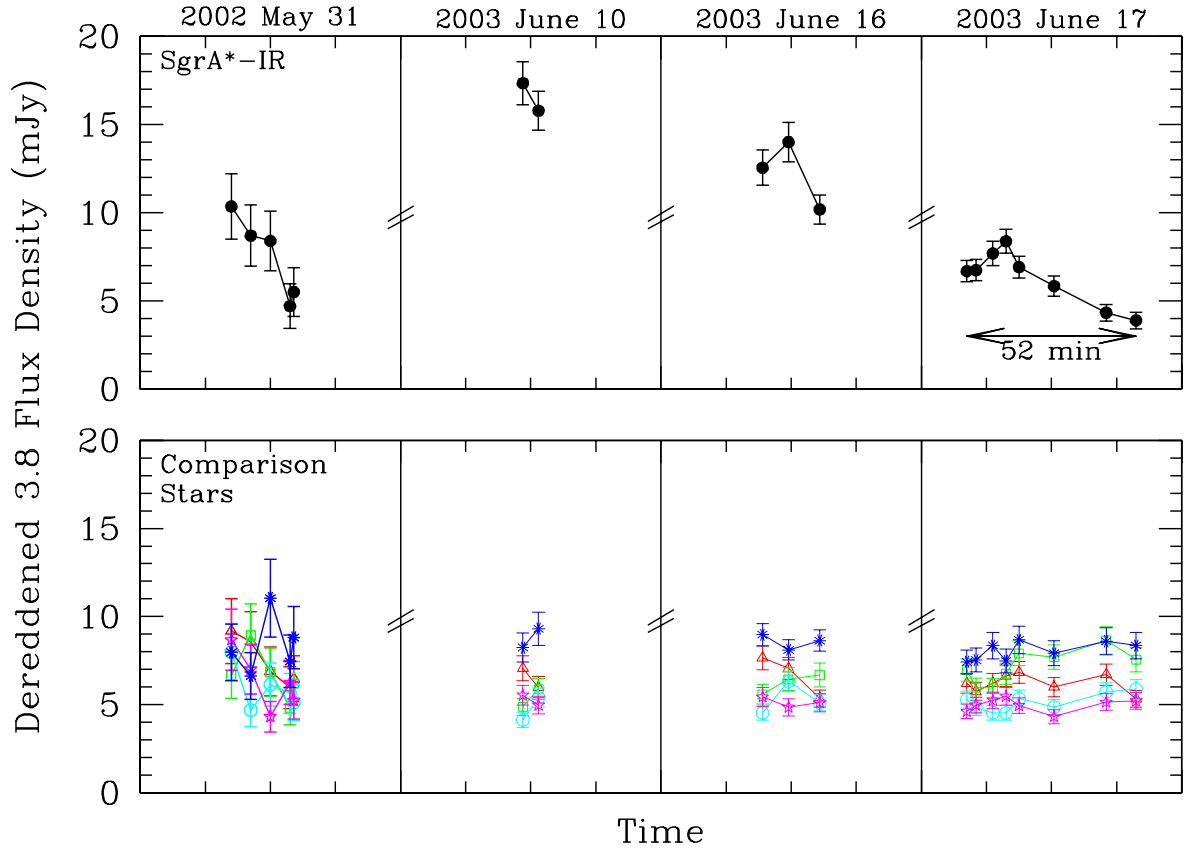}
\figcaption{Dereddened 3.8 $\mu$m flux densities for SgrA*-IR (top)
and 5 nearby stellar sources of similar brightness (bottom).  
A panel of 80 minutes for each of the four nights of observations is shown.
The comparison sources are S0-1 (cyan circles), S0-3 (red triangles),
S0-4 (green squares), S0-6 (blue asterisks), S0-11 (magenta stars).
While the stellar sources show no significant variation, SgrA*-IR
has varied by a factor of 4 over the course of a week, 2003 June 10-17,
and a factor of 2 within 40 minutes on the night of 2003 June 17.
\label{fig_phot}
}
\end{figure}

The newly identified L' source has several properties that 
indicate that it is affiliated with the SMBH.
First, its location on the plane of the sky is coincident with the 
SMBH to within 18 mas (1 $\sigma$).  Second, it appears to 
be a stationary source with an upper limit on its transverse motion of 
700 km s$^{-1}$, which 
differs from the proper motions of $>$6,000 km s$^{-1}$ for other sources 
that have come within 10 mas of the dynamical center (Ghez et al. 2003b;
Sch\"odel et al. 2003).  Third, it is a variable source with significant  
flux density changes observed on timescales as short as 40 min.
Fourth, it is an extremely red object.  In the K-band, a stationary source 
consistent with the location of the dynamical center would have to be
fainter than at least K $\sim$ 15.5 mag to be consistent with the observations 
of Ghez et al. (2003b) and Hornstein et al. (2002).  The K-L' color of the new 
source must therefore be greater than 2.1 mag, which is at least 1.0 mag redder 
than all the nearby stars used
as comparison stars in Figure~\ref{fig_phot}.  Such a red color is expected
for Sgr A*
from the sub-mm detections (e.g., Falcke et al. 1998; Zhao et al. 2003)
and 2.2 $\mu m$ limits (Hornstein et al. 2002) shown in Figure 3.
We conclude
that the new L' source is associated with the SMBH or, 
equivalently, Sgr A*, and therefore refer to it as SgrA*-IR. 

Several mechanisms associated with a SMBH can give rise
to infrared variability, including gravitational lensing (e.g., 
Alexander \& Loeb 2001), 
disk illumination (Cuadra, Nayakshin, \& Sunyaev 2003), 
star-disk collisions (Nayakshin \& Sunyaev 2003), and 
physical processes 
in the SMBH's accretion flow (e.g., Markoff et al. 2001; Liu \& Melia 2002;
Yuan, Quataert, \& Narayan 2003a,b).  Assuming that we are witnessing a single
phenomenon, the observed short timescales 
for the variations, $\sim$40 min, rule out the first two possibilities and
the longer timescales eliminate the third.  We therefore suggest
that the detected 3.8 $\mu$m emission emanates from either an accretion flow
or an outflow quite close to the SMBH, within 5 AU, or equivalently 80 
R$_{s}$.  While the shortest 
timescale variations observed at 3.8 $\mu$m are comparable
to the X-ray flare timescales (Baganoff et al. 2001, 2003b; 
Goldwurm et al. 2003; Porquet et al. 2003), the probability that
such an X-ray flare occurred during any of these observations 
is less than 
0.05 and the probability that such X-ray flares occurred or were 
in progress during all of our
observations is less than 10$^{-7}$.  
Nonetheless, the similarity in the timescales for variation at 3.8 $\mu$m 
and at X-ray wavelengths
lead us to suggest that 
the variable 3.8 $\mu$m emission is produced by a mechanism related to that 
which produces the X-ray flares.  We presume any associated variable X-ray 
emission 
during our observations  
was below the quiescent level. 

\begin{figure}
\epsscale{1.0}
\plotone{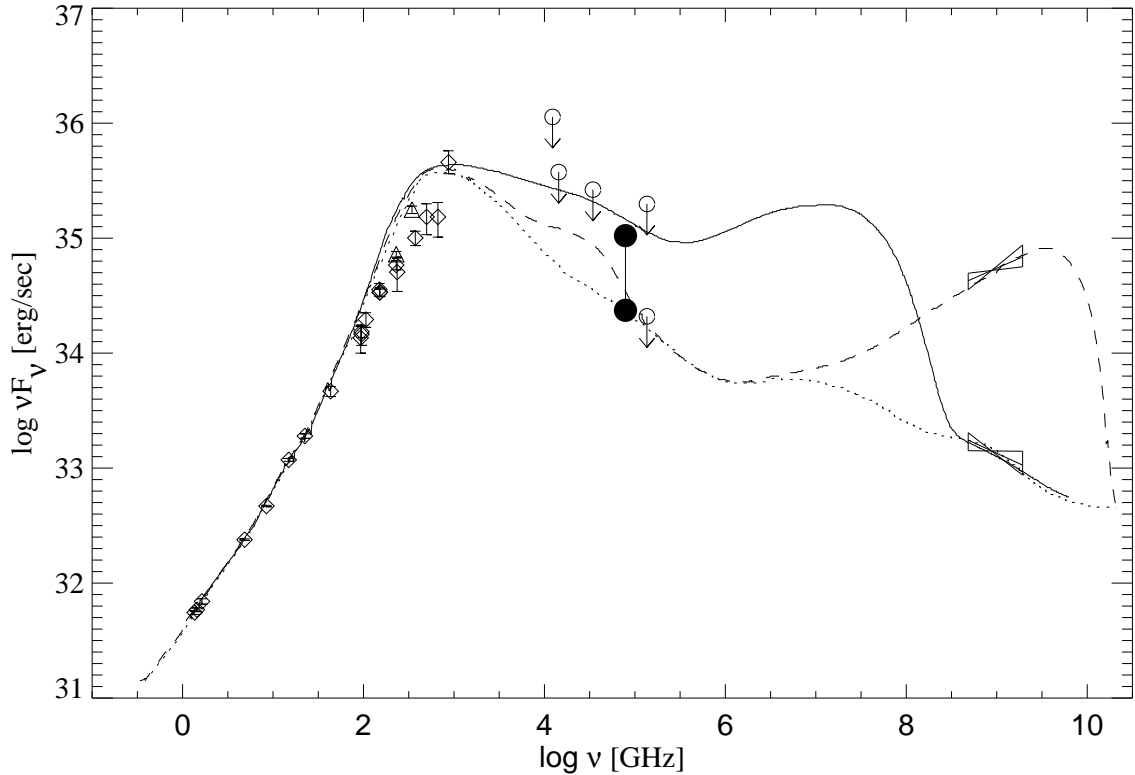}
\figcaption{The spectral energy distribution for Sgr A*.  The new 3.8 $\mu$m
flux density measurements are plotted as two filled circles delimiting the
range of observed values.  Other measurements from the 
literature are plotted with the following symbols: radio data appear 
as diamonds (Falcke et al. 1998)  and triangles (Zhao et al. 2003), mid-infrared
limits (Serabyn et al. 1997) and 2 $\mu$m limits (Hornstein et al. 2002)
are depicted with arrows, and X-ray flux densities in
steady state and a flaring state are plotted as bowties (Baganoff et al.
2001; 2003).  The curves show examples of models for the quiescent (dotted) 
and variable components (dashed - SSC model; solid - synchrotron only model) 
of Sgr A* taken 
from Yuan et al. (2003a,b).
\label{fig_sed}
}
\end{figure}

Several models of the flared X-ray emission have been proposed, invoking
physical processes such as elevated accretion rates or
enhanced electron acceleration from  
MHD turbulence, reconnection, or weak shocks 
and emission mechanisms
including bremsstrahlung, synchrotron, and inverse Compton
(Markoff et al. 2001; Liu \& Melia 2001, 2002; Yuan et al. 2003a,b).  
Models that account for the low variability amplitude at radio
frequencies generally invoke acceleration of some
fraction of the electrons into the power-law tail 
of the electron energy distribution. 
In these models, the X-ray flares arise from
either synchrotron emission, if a small fraction of the electrons are 
energetic enough ($\gamma \sim 10^6$), or
by synchrotron-self Compton (SSC) upscattering of sub-mm and infrared
photons, for a larger but less dramatically accelerated population 
of electrons.
In both cases, 
a significant amount of 3.8 $\mu$m emission can be produced 
from the direct synchrotron emission of the accelerated population
of electrons
(see Figure~\ref{fig_sed})
and the variability
may be induced by small changes in the fraction of electrons in the 
high-energy power-law tail of the electron energy distribution or by changes
in the value of the power-law (Yuan et al. 2003a).   
For the two cases that produce 3.8 $\mu$m emission via synchrotron emission,
the SSC model predicts correlated IR and X-ray
flares, while the synchrotron only model can more readily allow for an 
IR only flare, depending on the spectrum of the accelerated electron energy 
distribution.  It therefore appears at present that the synchrotron only
model best fits the available data 
(e.g., Yuan et al. 2003b).
Furthermore, in contrast to the X-ray flares, which suggest that 
processes giving rise to a high-energy power-law tail in the electron 
energy distribution occur infrequently ($\gtrsim$2\% of the time), 
the persistently detected
variable 3.8 $\mu$m emission suggests that such processes
are in effect much more often.
More generally, the 3.8 $\mu$m emission 
opens up a continuously accessible window for studying the conditions
and physical processes within the plasma that is either 
accreting onto or outflowing from the central black hole.

\acknowledgements

The authors thank J. Aycock, R. Campbell, 
G. Hill, C. Sorensen, M. van Dam, \& C. Wilburn
at the Keck Observatory for their help in obtaining the observations
\& G. Duch\^ene, J. Lu, F. Melia, R. Narayan, F. Yuan,
\& an anonymous referee for useful comments.
Support for this work was provided by NSF
grant AST-9988397 \& the NSF Science
\& Technology Center for AO, managed by UCSC 
(AST-9876783).
DT \& BTS are supported by the SIRTF project. SIRTF is carried
out at the JPL, operated by the Caltech, under contract with NASA.
The Keck Observatory is operated as a scientific partnership among 
Caltech, the University of California, \&
NASA.  The Observatory was made
possible by the generous financial support of the Keck Foundation.

\pagebreak

\pagebreak

\begin{deluxetable}{lllrrr}
\tablewidth{0pt}
\tablecaption{Summary of SgrA*-IR's 3.8 $\mu$m properties\label{tab_sum}}
\tablehead{\colhead{UT Date}&
	\colhead{UT Time}  &
	\colhead{L'}  &
	\colhead{F$_{\nu, dereddenned}$ }  &
	\multicolumn{2}{c}{Offset From Dynamical Center} \\
  	\colhead{} &
  	\colhead{} &
  	\colhead{(mag)} &
	\colhead{(mJy)} &
	\colhead{$\Delta$RA (mas)} &
	\colhead{$\Delta$Dec (mas)} 
}
\startdata
2002 May 31& Average	& 13.23 $\pm$ 0.10 & 6.9 $\pm$ 0.7 	& 0 $\pm$ $~$6 & -7 $\pm$  $~$7 \\
	    &   11:29		& 12.78 $\pm$ 0.18 & 10.35 $\pm$ 1.9 & \nodata & \nodata \\
	    &   11:35		& 12.97 $\pm$ 0.20 & 8.7 $\pm$ 1.7 & \nodata & \nodata \\
	    &   11:41		& 13.01 $\pm$ 0.20 & 8.4 $\pm$ 1.7 & \nodata  & \nodata \\
	    &   11:47		& 13.64 $\pm$ 0.27 & 4.7 $\pm$ 1.3 & \nodata  & \nodata \\
	    &   11:52		& 13.47 $\pm$ 0.25 & 5.5 $\pm$ 1.4 & \nodata & \nodata \\
2003 Jun 10 &	Average\tablenotemark{c}		& 12.28	$\pm$ 0.05 & 16.4 $\pm$ 0.8  & -8 $\pm$ $~$9  & -1 $\pm$ 10 \\
	    &   12:06        	& 12.22 $\pm$ 0.07 & 17.4 $\pm$ 1.2 & \nodata & \nodata \\
	    &   12:10 		& 12.32 $\pm$ 0.07 & 15.8 $\pm$ 1.1 & \nodata & \nodata \\
2003 Jun 16 &	Average\tablenotemark{c}		& 12.54	$\pm$ 0.05 & 12.9 $\pm$ 0.6  & -15 $\pm$ $~$9 & 9 $\pm$ $~$8 \\
	    &   11:20		& 12.56 $\pm$ 0.08 & 12.6 $\pm$ 1.0 & \nodata & \nodata \\
	    &   11:28           & 12.45 $\pm$ 0.08 & 14.0 $\pm$ 1.1 & \nodata & \nodata \\
	    &   11:38 		& 12.80	$\pm$ 0.09 & 10.2 $\pm$ 0.9 & \nodata & \nodata \\
2003 Jun 17 &	Average\tablenotemark{c}		& 13.38	$\pm$ 0.04 & 5.9 $\pm$ 0.2 & -12 $\pm$ 13  & -7 $\pm$ 18 \\
	    &	10:32		& 13.25 $\pm$ 0.09 & 6.7 $\pm$ 0.6 & \nodata & \nodata \\
	    &	10:36		& 13.24 $\pm$ 0.09 & 6.7 $\pm$ 0.6 & \nodata & \nodata \\
	    & 	10:41		& 13.10 $\pm$ 0.09 & 7.7 $\pm$ 0.7 & \nodata & \nodata \\
	    & 	10:45		& 13.01 $\pm$ 0.08 & 8.4 $\pm$ 0.7 & \nodata & \nodata \\
	    &	10:49		& 13.22 $\pm$ 0.09 & 6.9 $\pm$ 0.6 & \nodata & \nodata \\
	    &	10:59		& 13.40 $\pm$ 0.10 & 5.8 $\pm$ 0.6 & \nodata & \nodata \\
	    &	11:16		& 13.72 $\pm$ 0.11 & 4.3 $\pm$ 0.5 & \nodata & \nodata \\ 
	    &	11:24		& 13.84 $\pm$ 0.12 & 3.9 $\pm$ 0.5 & \nodata & \nodata \\

\enddata
\tablenotetext{a}{The reported photometric uncertainties incorporate only
uncertainties in the relative photometry and are based on a scaling of
the reference stars' average rms ($<$L'$>$$\sim$13.2 mag) each night.}
\tablenotetext{b}{Since the new L' source coincided with S0-2 in 2002, the report
ed magnitude
has had the flux density observed for S0-2 in 2003 removed (see \S4).}
\tablenotetext{c}{The average values are from an analysis of the maps that
are averages of all the selected frames in a given night (see \S3)}
\end{deluxetable}

\pagebreak

\pagebreak

\pagebreak

\end{document}